\documentstyle[aps,twocolumn,prl,epsfig,axodraw]{revtex}

\newcommand{\be}{\begin{equation}}
\newcommand{\ee}{\end{equation}}
\newcommand{\br}{\begin{eqnarray}}
\newcommand{\er}{\end{eqnarray}}
\newcommand{\ba}{\begin{array}}
\newcommand{\ea}{\end{array}}
\newcommand{\bi}{\begin{itemize}}
\newcommand{\ei}{\end{itemize}}
\newcommand{\bn}{\begin{enumerate}}
\newcommand{\en}{\end{enumerate}}
\newcommand{\bc}{\begin{center}}
\newcommand{\ec}{\end{center}}

\begin{document}
\twocolumn[\hsize\textwidth\columnwidth\hsize\csname @twocolumnfalse\endcsname


\title{Weak corrections and high $E_T$ jets at Tevatron}
\author{S. Moretti, M.R. Nolten and D.A. Ross}
\address{School of Physics \& Astronomy, University of Southampton,
Highfield, Southampton SO17 1BJ, UK} 

\maketitle
\begin{abstract}
\noindent
We calculate one-loop (purely) Weak (W) corrections of
${\cal O}(\alpha_{\mathrm{S}}^2\alpha_{\mathrm{W}})$ 
to the partonic cross section
of two jets  at Tevatron and prove that they can be  larger 
than the tree-level ${\cal O}(\alpha_{\mathrm{S}}\alpha_{\mathrm{EW}})$ 
and ${\cal O}(\alpha_{\mathrm{EW}}^2)$ Electro-Weak (EW)
ones. At high transverse energy of the jets, all such corrections
may lead to detectable effects of, e.g., $-10\%$ or so, with respect to the leading-order (LO)
QCD term of  ${\cal O}(\alpha_{\mathrm{S}}^2)$, for the highest value so far probed by Run 2, 
depending on the factorisation/renormalisation scale. Besides, they increase
significantly with jet transverse energy. Hence, our results show that EW corrections
may be needed to fit the Standard Model (SM) to present and future Tevatron 
jet data. 
\end{abstract}

\vspace*{3mm}
]

\noindent
As the overall energy of hard scattering processes increases one should
expect a  relatively large impact of
perturbative  EW corrections, as
compared to the QCD ones. This can easily be understood 
(see \cite{Melles:2001ye}~--\cite{Denner:2001mn} and references therein
for reviews) in terms of the so-called
Sudakov (leading) logarithms of the form 
$\alpha_{\mathrm{W}}\log^2({\hat{s}}/M_{W}^2)$
(hereafter, $\alpha_{\rm{W}}\equiv\alpha_{\rm{\small EM}}/\sin^2\theta_{\rm W}$,
with $\alpha_{\rm{\small EM}}$ the Electro-Magnetic (EM) coupling constant and
$\theta_{\rm W}$ the weak mixing angle, whereas $\sqrt{\hat{s}}$ is
the parton-level centre-of-mass energy), which appear
in the presence of higher order weak corrections when the initial state
carries a definite non-Abelian flavour and which, unlike QCD,
do not cancel between virtual and real emission 
of $W$ bosons \cite{Ciafaloni:2000df-Ciafaloni:2000rp}.

Furthermore, one should recall that real weak bosons are unstable and 
decay into high
transverse momentum leptons and/or jets, which are normally
captured by the detectors. In the definition of a hadronic cross section,
one may then remove events with such additional particles.
 Hence, for typical experimental
resolutions, softly and collinearly emitted weak bosons need not be included
in the definition of the
production cross section and one can restrict oneself to the calculation
of weak effects originating from virtual corrections only. In fact,
leading (and all subleading) virtual weak corrections are finite
(unlike QCD, where infrared divergences mean that virtual corrections must be 
considered in conjunction with gluon bremsstrahlung), 
as the mass of the weak gauge boson provides a physical
cut-off for the otherwise divergent infrared behaviour.
Under these circumstances,
the (virtual) exchange of $Z$ bosons also generates double logarithmic
corrections, 
$\alpha_{\mathrm{W}}\log^2({\hat{s}}/M_{Z}^2)$.
Moreover, in some simpler cases, the genuinely weak contributions can  be
isolated in a gauge-invariant manner from purely EM effects
and the latter may or may not
be included in the calculation, depending on the observable being studied. 

The leading, double-logarithmic, angular-independent weak
logarithmic corrections are universal, i.e., they depend only on the identities
of the external particles. 
In some instances, however, large cancellations between angular-independent
and angular-dependent corrections \cite{Accomando:2001fn}
(see also \cite{Denner:2003wi} for two-loop results)
 and between leading and
subleading terms \cite{Kuhn:2001hz} have been found at TeV energies.
Moreover, some other considerations are in order in the specific hadronic
context. Firstly, one should recall that hadron-hadron scattering events 
involve valence (or sea) partons of opposite isospin in the same process, 
but since the PDFs are not singlets of flavour
only partial cancellations among initial state large logarithms will occur
\cite{Ciafaloni:2000df-Ciafaloni:2000rp}.
Secondly, several crossing symmetries among the involved partonic subprocesses
can also easily lead to more cancellations. 

Because all this,  it becomes of crucial importance 
to study the full set of fixed order
weak corrections, in view of establishing the relative size of the different
contributions at the energies which can be probed at TeV scale
hadronic machines. Several results already exist, e.g., in the SM, for: 
EW gauge boson production in single mode
\cite{Accomando:2001fn},  
\cite{Baur:1998kt-Kuhn:2004em} as well as in pairs
\cite{Pozzorini:2001rs-Hollik:2004tm}; 
$b\bar b$ 
\cite{Maina:2003is} and
$t\bar t$ \cite{Kuhn:2005it}~--\cite{Kao:1999kj} production;
Higgs processes 
\cite{Ciccolini:2003jy}. 
(See \cite{Hollik:2004dz} for a review.)

It is the aim of our paper to report on the computation of the full
one-loop weak effects\footnote{We neglect considering here
purely EM effects (as well as interferences between these and the
weak ones), 
as they can be isolated in a gauge invariant
fashion and since they are not associated with logarithmic
enhancements either (like QCD).} entering all possible
`2 parton $\to$ 2 parton' scatterings, through the
perturbative order $\alpha_{\mathrm{S}}^2\alpha_{\mathrm{W}}$. 
(See Ref.~\cite{Baur:1989qt} for
tree-level $\alpha_{\mathrm{S}}\alpha_{\mathrm{EW}}$ interference
effects -- hereafter,
$\alpha_{\mathrm{EW}}$ exemplifies the fact that
both EM and W effects are included at the given order).
We will ignore altogether the contributions
of tree-level $\alpha_{\mathrm{S}}^2\alpha_{\mathrm{W}}$ terms
involving the radiation of $W$ and $Z$ bosons. Therefore,
apart from $gg\to gg$, $qq'\to QQ'$, $\bar q\bar q'\to \bar Q\bar Q'$ and  $q\bar q'\to Q\bar Q'$
(which are not subject to order 
$\alpha_{\mathrm{S}}^2\alpha_{\mathrm{W}}$ corrections), 
there are in total fifteen subprocesses to consider,
\begin{eqnarray}
g g &\to& q \bar q,\qquad\qquad\qquad\qquad\qquad q \bar q \to g g,\\ 
q g &\to& q g,\qquad\qquad\qquad\qquad\qquad\bar q g \to \bar q g,\\ 
q q &\to& q q,\qquad\qquad\qquad\qquad\qquad \bar q \bar q \to \bar q \bar q,\\
q Q &\to& q Q ~({\rm{same~or~different~generation}}),\\
\bar q \bar Q &\to& \bar q \bar Q ~({\rm{same~or~different~generation}}),\\
q \bar q &\to& q \bar q, \\
q \bar q &\to& Q \bar Q ~({\rm{same~or~different~generation}}),\\
q \bar Q &\to& q \bar Q ~({\rm{same~or~different~generation}}),
\end{eqnarray}
with $q^{(')}$ and $Q^{(')}$ referring to quarks of different flavours,
limited to $u$-, $d$-, $s$-, $c$- 
and $b$-type (all massless). While the first four
processes (with external gluons) were already computed 
in Ref.~\cite{Ellis:2001ba}, 
the eleven four-quark processes are new to this study
(see Ref.~\cite{Moretti:2005aa} for
RHIC and LHC results).
Besides, unlike the channels with external gluons, those with four-quarks
must include virtual gluon corrections to tree-level interferences between
weak and strong interactions and therefore 
can be infrared
divergent, which means that
gluon bremsstrahlung effects must be evaluated to obtain
a finite cross section at the given order. In addition,
for completeness, we have  included 
the non-divergent subprocesses of (anti-)quark-gluon scattering
into three coloured fermions.

Our studies are of particular relevance in the context
of the Tevatron collider, where an excess was initially found
by CDF (but not D0)  at high transverse energy in inclusive jet 
data from Run 1 \cite{Affolder:2001fa}, 
with respect to the next-to-LO (NLO) QCD predictions 
\cite{Aversa:1988fv}~--\cite{Giele:1994gf}. 
While several
speculations were made about the possible
sources of such excess from physics beyond
the SM, it was eventually
pointed out that a modification of the
gluon PDFs at medium/large Bjorken $x$ can apparently
reconcile theory and data within current
systematics: see, e.g., \cite{Stump:2003yu}.
(For a different explanation, see \cite{Martin:2004ir-Klasen:1996yk}.)
In fact, notice that with the most recent PDFs 
(e.g., CTEQ6.1M \cite{Pumplin:2002vw}), also preliminary Run 2
 data seem to be (barely) consistent with NLO QCD,
see \cite{newexcess} for CDF.
(Results from D0 have a larger systematic uncertainty, which tends to 
encompass the theory predictions \cite{newexcess}.)

Over a hundred one-loop and tree-level 
diagrams are involved in the computation of processes (1)--(8)
and is thus of paramount
importance to perform careful checks. In this respect, we 
should mention that our  expressions 
have been calculated independently
by at least two of us using FORM \cite{Vermaseren:2000nd} and 
that some results have also been reproduced by another
program based on FeynCalc \cite{Kublbeck:1990xc}. 

As already mentioned, infrared divergences  occur when the virtual or 
real (bremsstrahlung) gluon is either soft or collinear with the emitting parton
and these have been dealt with by using the formalism
of Ref.~\cite{Catani:1996vz}, whereby corresponding dipole terms 
are subtracted from the bremsstrahlung contributions in order to render the
 phase space integral free of infrared divergences. The
 integration over the gluon phase space of these dipole terms
 was performed analytically in $d$-dimensions, yielding pole terms
which cancelled explicitly against the pole terms of the virtual graphs.
There remains a divergence from the initial state collinear configuration,
which is absorbed into the scale dependence of the PDFs and must be matched
to the scale at which these PDFs are extracted. Through the order at which we are
working, it is sufficient to take the LO evolution of the PDFs 
(and thus the one-loop running of $\alpha_{\rm{S}}$). 

Some of the diagrams also  contain ultraviolet divergences.
These have been subtracted using the `modified' Dimensional Reduction
(${\overline{\mathrm{DR}}}$) scheme at
the scale $\mu=M_Z$. The use of ${\overline{\mathrm{DR}}}$,
as opposed to the more usual `modified' Minimal Subtraction
(${\overline{\mathrm{MS}}}$) scheme, is forced
upon us by the fact that the $W$- and $Z$-bosons contain axial couplings
which cannot be consistently treated in ordinary dimensional 
regularisation. 
Although not essential, we find it
convenient to work with helicity matrix elements extracted using properties
of Dirac matrices valid in four dimensions. 
 Thus the values taken for $\alpha_{\rm S}$ refer to the
${\overline{\mathrm{DR}}}$ scheme whereas the EM coupling,
$\alpha_{\mathrm{EM}}$, 
has been taken to be $1/128$ at the above subtraction
point.  (The numerical difference between these two schemes is
negligible for $\alpha_{\rm S}$ though.)

For the top mass and width, entering some of the loop diagrams with external
$b$-quarks, we have taken $m_t=175$ GeV and $\Gamma_t=1.55$ GeV, respectively.
The $Z$ mass used was $M_Z=91.19$ GeV and was related to the $W$ mass, $M_W$, via the
SM formula $M_W=M_Z\cos\theta_W$, where $\sin^2\theta_W=0.232$.
(Corresponding widths were $\Gamma_Z=2.5$ GeV and $\Gamma_W=2.08$ GeV.)

\begin{figure}[h!]
\hspace*{0.005truecm}{\epsfig{file=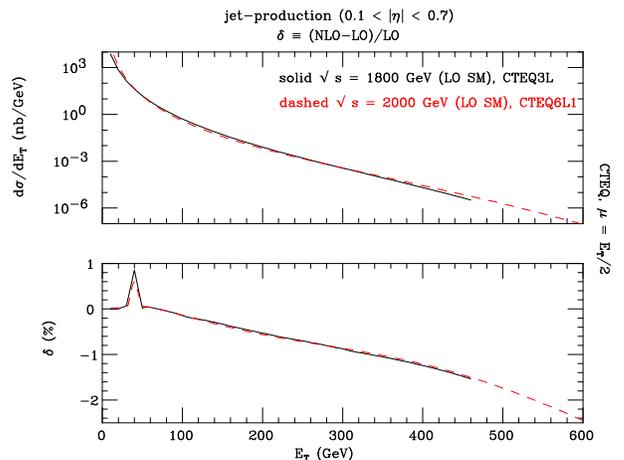,height=80mm,angle=90}}
\caption{The  effects of the 
${\cal O}(\alpha_{\mathrm{S}}^2\alpha_{\mathrm{W}})$
corrections relative to the LO SM results
for Run 1(Run 2) using CTEQ3L(CTEQ6L1) as  PDFs. 
}
\label{fig:TevatronExcesses}
\end{figure}

Fig.~\ref{fig:TevatronExcesses} shows the effects of our one-loop corrections
to the LO results for jet production, the latter being defined
as including all possible terms of order
$\alpha_{\mathrm{S}}^2$, 
$\alpha_{\mathrm{S}}\alpha_{\mathrm{EW}}$ and
$\alpha_{\mathrm{EW}}^2$ (hereafter LO SM). (The spike at $E_T\approx M_W/2,
M_Z/2$ is a threshold effect in the loop diagrams.)
Notice that in our treatment we identify the jets with the partons
from which they originate and we adopt here the cut $0.1<|\eta|<0.7$
in pseudorapidity to mimic the CDF detector coverage and
the standard jet cone requirement $\Delta R>0.7$ to emulate the
jet data selection (although we eventually sum the two- and three-jet contributions). Furthermore,
as factorisation and renormalisation scale we use $\mu=\mu_F\equiv\mu_R=E_T/2$ -- 
a choice leading to the best convergence of both NLO \cite{Giele:1994gf} and resummed  
\cite{Kidonakis:2000gi} QCD predictions  --
(where $E_T$ is the jet transverse
energy) while
we adopt 
CTEQ3L as PDFs \cite{Pumplin:2002vw} for Run 1, a set defined prior to the re-arrangement
of the gluon. With respect to the LO SM rates, the 
 ${\cal O}(\alpha_{\mathrm{S}}^2\alpha_{\mathrm{W}})$ corrections are not large
despite growing steadily with $E_T$. For $E_T$ values in the
vicinity of 420 GeV, the highest point of Run 1 and
also the location of the apparent CDF excess, they amount to $-1.5\%$.
This effect is not competitive with the positive NLO QCD corrections
through ${\cal O}(\alpha_{\rm{S}}^3)$: see, e.g., 
Fig.~1 of \cite{Giele:1994gf}. In the same figure, we have also shown the
 ${\cal O}(\alpha_{\mathrm{S}}^2\alpha_{\mathrm{W}})$  
corrections at Run 2 for the same $\mu$ and the choice  CTEQ6L1 of PDFs 
(one of the newest sets incorporating the above mentioned gluon
re-parameterisation). Here, we have also increased the $E_T$
values probed, as the larger collider energy has already allowed to collect
data some 150 GeV beyond the Run 1 reach. We see that at the higher energy the
${\cal O}(\alpha_{\mathrm{S}}^2\alpha_{\mathrm{W}})$ corrections are substantially
similar in size and shape to the lower energy case, so that they 
stretch to $-2\%$ or so near the current kinematic limit (550 GeV or so). 
(Crossing points between the two curves are induced by the different PDF choice
as well as the different numerical value of $\mu$ at the two energies.) 

\begin{figure}[h!]
\hspace*{0.005truecm}{\epsfig{file=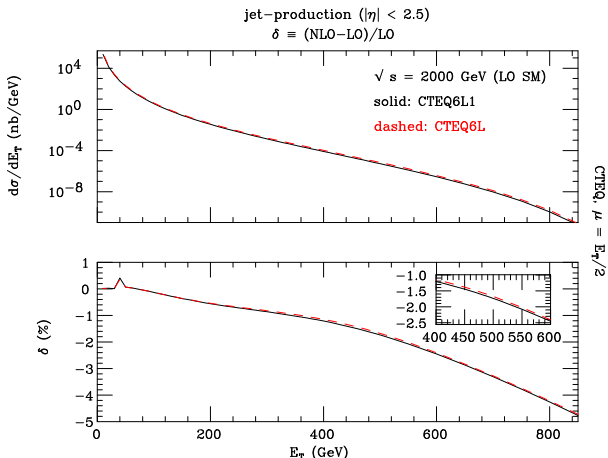,height=80mm,angle=90}}
\caption{The effects of the 
${\cal O}(\alpha_{\mathrm{S}}^2\alpha_{\mathrm{W}})$
corrections relative to the LO SM results 
for Run 2 in the presence of two sets of up-to-date PDFs (CTEQ6L and CTEQ6L1). 
}
\label{fig:Tevatron}
\end{figure}

Fig.~\ref{fig:Tevatron} extends the $E_T$ interval to 850 GeV and the
pseudorapidity cover to $|\eta|<2.5$ (our new default
from now on, for same $\Delta R$), while still adopting 
 $\mu=E_T/2$ as factorisation/renormalisation scale. 
Including the forward/backward detector region reduces minimally
the effects of the ${\cal O}(\alpha_{\mathrm{S}}^2\alpha_{\mathrm{W}})$ 
corrections while their shape remains unchanged. Their maximum is
about $-5\%$ at the upper end of the interval considered.
Furthermore, their dependence 
on the choice of PDFs is also very small,  as we have
verified by running CTEQ6L1 vs. CTEQ6L \cite{Pumplin:2002vw}.
  
\begin{figure}[h!]
\hspace*{0.005truecm}{\epsfig{file=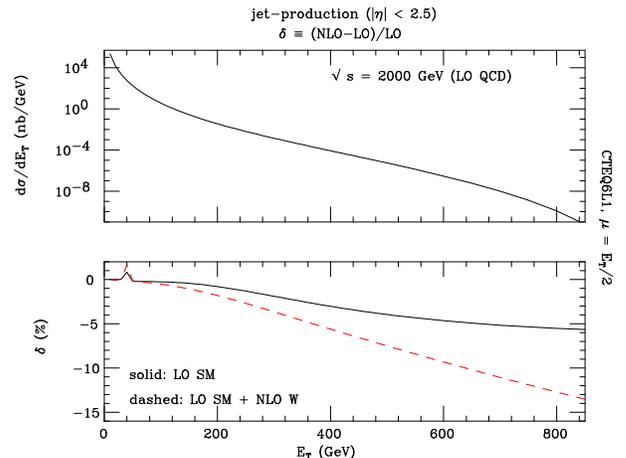,height=80mm,angle=90}}
\caption{The effects of the 
${\cal O}(\alpha_{\mathrm{S}}\alpha_{\mathrm{EW}}+\alpha_{\mathrm{EW}}^2)$ (LO SM)
and the latter plus
${\cal O}(\alpha_{\mathrm{S}}^2\alpha_{\mathrm{W}})$ (LO SM + NLO W)
corrections relative to the LO QCD of ${\cal O}(\alpha_{\mathrm{S}}^2)$
results
for Run 2 in the presence of up-to-date PDFs (CTEQ6L1). 
}
\label{fig:TevatronCorr}
\end{figure}

Notice however that, if one defines the corrections 
with respect to only the 
${\cal O}(\alpha_{\mathrm{S}}^2)$ contribution (hereafter, LO QCD),
the effects of the sum of all non-QCD terms, i.e., those of order 
$\alpha_{\mathrm{S}}\alpha_{\mathrm{EW}}$, 
$\alpha_{\mathrm{EW}}^2$ and $\alpha_{\mathrm{S}}^2\alpha_{\mathrm{W}}$ 
(hereafter LO SM + NLO W), become
 significantly larger. Fig.~\ref{fig:TevatronCorr} makes this
point clear. At $E_T=850$ GeV or so, the upper kinematic limit of the collider,
one would see a combined effect of about $-14\%$, most of which are indeed due to the 
${\cal O}(\alpha_{\mathrm{S}}^2\alpha_{\mathrm{W}})$
terms new to this study (NLO W). In practice though, 
such jet transverse energies are unreachable even for optimistic
luminosity. For the current Run 2 highest $E_T$ point, 550 GeV, 
the effects of the  LO SM + NLO W corrections amount to $-8\%$
of the LO QCD term. Clearly, it is of paramount importance to establish
which terms are included in Monte Carlo (MC) programs used to
interpolate the data. In general, it is clear from Fig.~\ref{fig:TevatronCorr}
that the corrections due to the one-loop graphs play a role at least as relevant as those
due to tree-level effects and, importantly, at Tevatron, they act in the same direction, 
namely, a reduction of the differential QCD rates.

\begin{figure}[h!]
\hspace*{0.005truecm}{\epsfig{file=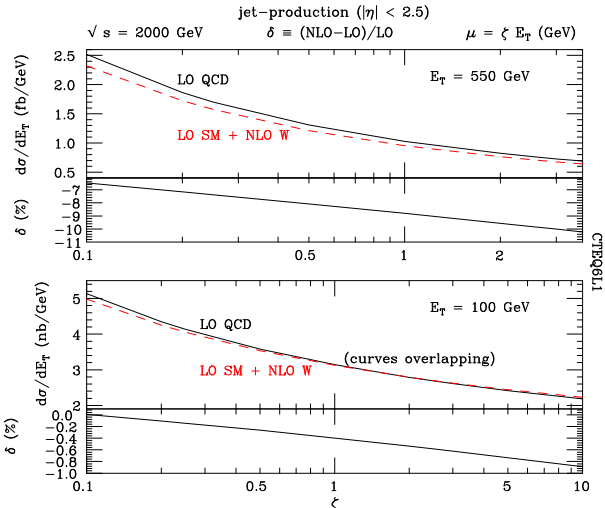,height=80mm,angle=90}}
\caption{The effects of the 
${\cal O}(\alpha_{\mathrm{S}}\alpha_{\mathrm{EW}}+\alpha_{\mathrm{EW}}^2
+\alpha_{\mathrm{S}}^2\alpha_{\mathrm{W}})$ (LO SM + NLO W)
corrections relative to the LO QCD
results as function of $\mu$
for Run 2 in the presence of up-to-date PDFs (CTEQ6L1)
for two choices of jet transverse energy. 
}
\label{fig:TevatronRescale}
\end{figure}

In fact, another subtlety should be borne in mind as far as EW corrections
are concerned. We have so far adopted $\mu=E_T/2$ for the factorisation/renormalisation
scale. This seems in fact to be the preferred choice while
comparing Tevatron data against NLO QCD predictions through 
${\cal O}(\alpha_{\rm{S}}^3)$.
A discussion of the dependence of the QCD corrections on $\mu$
is found in Refs.~\cite{Ellis:1990ek}~--\cite{Giele:1994gf}
and the above mentioned choice is motivated by the stability
of the higher order QCD results in the region $\mu\approx E_T/2$. In fact, recall that any 
dependence on $\mu$ arises because of the truncation of the perturbative 
expansion at some fixed order and it is therefore a measure of the missing 
higher order terms. As $\mu$ would not appear if these were known through all orders, 
it is customary to vary the factorisation/renormalisation scale in order to 
estimate the residual theoretical error. 
We have done so in Fig.~\ref{fig:TevatronRescale} for, 
e.g., $E_T=100$ and $550$ GeV, at Run 2 energy with CTEQ6L1
as PDFs. The fact that the 
${\cal O}(\alpha_{\mathrm{S}}^2\alpha_{\mathrm{W}})$ curves 
do not display local maxima, unlike the $
{\cal O}(\alpha_{\mathrm{S}}^3)$ results
(Fig.~2 of \cite{Ellis:1990ek}), does intimate that one scale choice
is not more appropriate than another
(irrespective of the jet transverse energy probed and the size
of the EW corrections). Thus, there is no firm reason to adopt $E_T/2$ 
as factorisation/renormalisation scale
here. If a higher value is chosen at 550 GeV, e.g., $\mu=E_T$, 
the LO SM + NLO W corrections 
grow of a further percent, to $-9\%$, while for $\mu=2E_T$ they become $-10\%$. This trend
is manifest over the entire $E_T$ range of relevance at Tevatron. 

In summary, at Tevatron, EW effects in general and 
${\cal O}(\alpha_{\mathrm{S}}^2\alpha_{\mathrm{W}})$ 
one-loop terms in particular are
important contributions to the inclusive jet cross section at
large transverse energy. A careful re-analysis of actual jet data, which
was beyond the intention of this paper, may be needed in view
of the increasing luminosity of the Fermilab collider. Particular care
should be devoted to the treatment of real $W$ and $Z$ production and
decay in the definition of the inclusive jet data sample, as this will determine
whether  tree-level $W$ and $Z$ bremsstrahlung effects have to be included
in the theoretical predictions through ${\cal O}(\alpha_{\mathrm{S}}^2\alpha_{\mathrm{W}})$,
which might counterbalance the negative effects due to  one-loop
$W$ and $Z$ virtual exchange.  In closing, we should mention
that the calculation of the aforementioned EM effects is in progress.

\end{document}